\documentclass[cits]{PoS}

\title{The physical and compositional properties of dust: what do we really know?}

\ShortTitle{The physical and compositional properties of dust}

\author{\speaker{Ant Jones} \\ 
        Institut d'Astrophysique Spatiale, CNRS/Universit\'e Paris Sud, Orsay, 91405, France \\
        E-mail: \email{Anthony.Jones@ias.u-psud.fr}}

\abstract{
Many things in current interstellar dust studies are taken as well understood givens by much of the community. For example, it is widely held that interstellar dust is made up of only three components, {\it i.e.}, ``astronomical silicates'', graphite and polycyclic aromatic hydrocarbons, and that our understanding of these is now complete and sufficient enough to interpret astronomical observations of dust in galaxies. 
To zeroth order this is a reasonable approximation. 
However, while these ``three pillars'' of dust modelling have been useful in advancing our understanding over the last few decades, it is now apparent that they are insufficient to explain the observed evolution of the dust properties from one region to another.
Thus, it is time to abandon the ``three pillars'' approach and to seek more physically-realistic interstellar dust analogues. 
The analysis of the pre-solar grains extracted from meteorites, interplanetary dust particles and from the Stardust mission, and the interpretation of x-ray scattering and absorption observations, supports the view that our current view of the interstellar dust composition(s) is indeed too na\"{i}ve. 
The aim of this review is to point out where our current views are rather secure and, perhaps more importantly, where they are far from secure and we must re-think our ideas. 
To this aim ten aspects of interstellar dust will be scrutinised and re-evaluated in terms of their validity within the current observational, experimental, modelling and theoretical constraints. 
It is concluded from this analysis that we really do need to re-assess many of the fundamental assumptions relating to what we think we really do `know' about interstellar dust. In particular, it is clear that unravelling the nature dust evolution in the interstellar medium is perhaps {\em the} key to significantly advancing our current understanding of interstellar dust. For example, the dust in the diffuse interstellar medium, molecular clouds, photo-dissociation regions and HII regions is not exactly the same but exhibits important evolution within and between these different regions. An understanding of these evolutionary and regional variations exhibited by dust is now critical. 
}

\FullConference{The Life Cycle of Dust in the Universe: Observations, Theory, and Laboratory Experiments - LCDU 2013,\\
		18-22 November 2013\\
		Taipei, Taiwan}

\begin{document}

\setcounter{section}{-1}
\section{Background}
\label{sect_intro}

Interstellar dust has been the subject of dedicated studies for more than 80 years, ever since the early measurements of interstellar reddening by Trumpler \cite{1930PASP...42..214T}. Dust models followed about a decade or so after these early dust extinction measurements and one of the earliest was the dirty ice model of van de Hulst \cite{vandehulst43}. Oort \& van de Hulst \cite{1946BAN....10..187O} then considered the processing of these dirty ice particles in the interstellar medium (ISM) and estimated a dust lifetime of about 50 million years.
Some thirty years later dust modelling became more sophisticated when Mathis {\it et al.} studied uncoated graphite, enstatite, olivine, silicon carbide, iron and magnetite particles as viable dust materials to explain extinction in the ISM \cite{1977ApJ...217..425M}. They concluded that graphite was a necessary dust component of any viable mixture and that it could be combined with any of the other materials to satisfactorily explain interstellar extinction. So were born our current ideas about interstellar dust being composed of graphite and some form of silicate. Thus, and for almost forty years, graphite and amorphous silicate materials have formed the basis of the most widely used dust models and they have served us well. Nevertheless, it is perhaps to time to re-visit our long-held views on the nature of dust in the ISM and to re-examine some of our basic assumptions. Thus:  

{\bf What do we know of dust?}  
Currently we believe that we have a pretty good understanding of the nature of interstellar dust in our own and also in distant galaxies. The dust is evidently made of ``astronomical'' amorphous silicates, graphite and polycyclic aromatic hydrocarbons (PAHs), and our understanding of these is now complete and sufficient enough to allow for a good interpretation of astronomical observations.  
This understanding has been sufficient to meet our needs for the last few decades but we are now finding clear and systematic differences in the dust from one region to another, differences that cannot be accounted for with the `standard' astronomical silicate, graphite and PAH model. 

{\bf What do we really know of dust?}
Thus, and upon closer examination of the `facts' it seems likely that in many, if not most, cases we need to review our current understanding, push aside the veil of complacency and to re-think our ideas. The major aim of this review is therefore to take a careful and critical look at some of our fundamental ideas about dust in space. 

{\bf Where do we go from here?}
One of the major conclusions of this close re-consideration is that in interstellar dust studies we really need to adopt more physically-realistic models, which means that  from here on the dust physics gets much more interesting, but also much more complex. 

In the following ten sections some aspects of the observable properties of dust and what we really know of cosmic dust are examined in some detail. 

\section{The UV-FUV extinction properties of dust}
\label{sect_dust_ext}

\begin{figure}[h] 
\includegraphics[angle=270.,width=.5\textwidth]{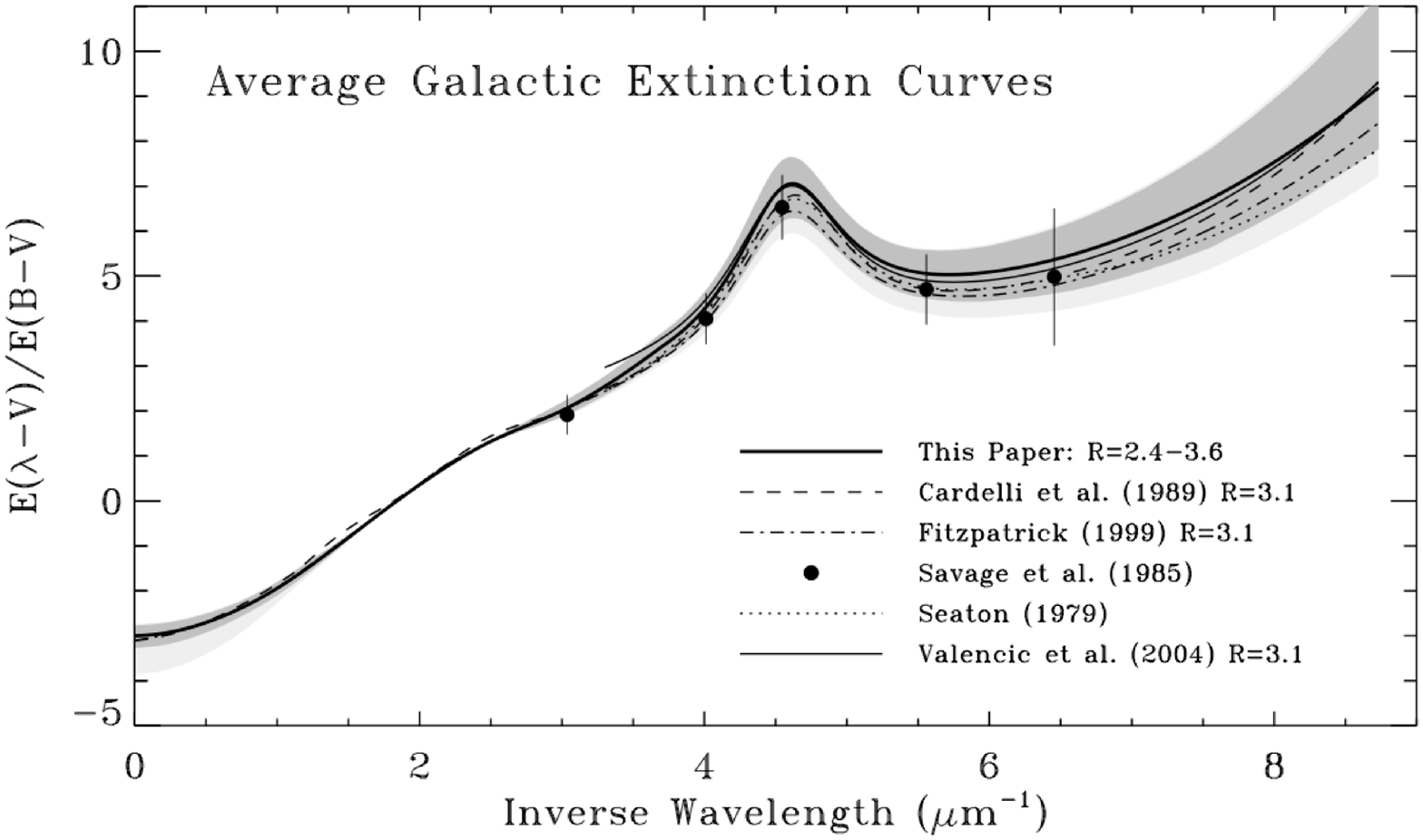} 
\includegraphics[angle=270.,width=.5\textwidth]{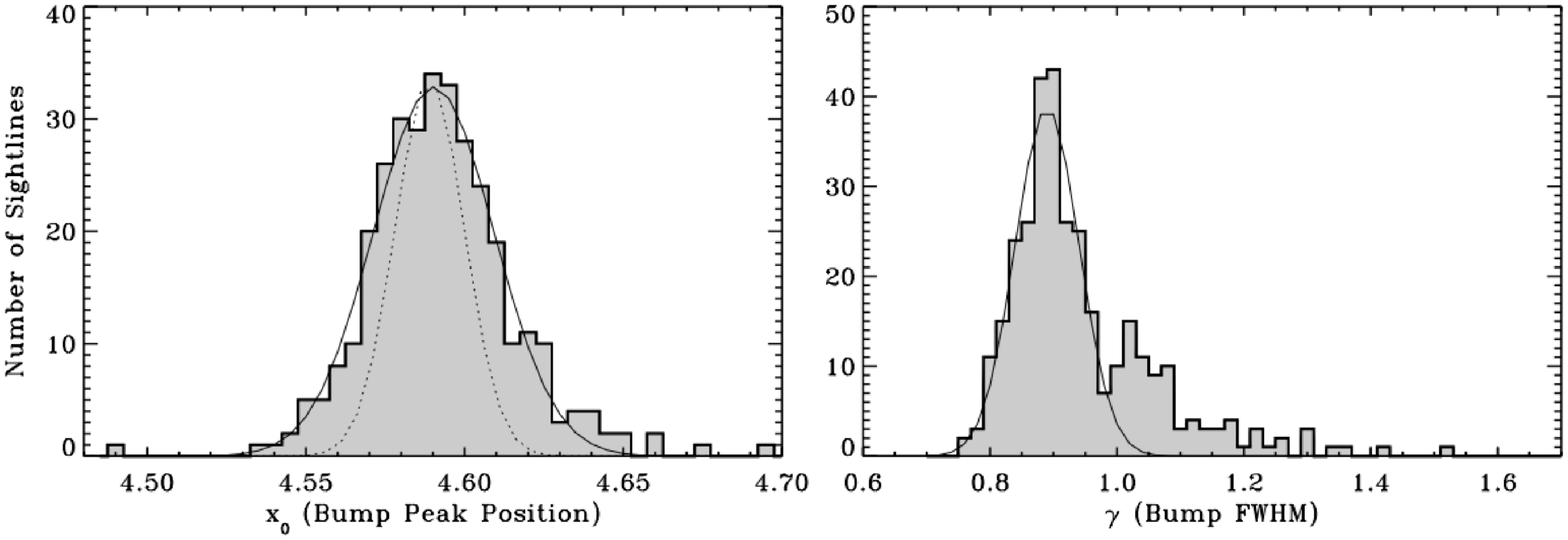} 
\caption{The $E(B-V)$-normalised average galactic extinction curves (left) and the observed UV extinction bump position and width variations (right): Figs. 9 and 17 from \cite{2007ApJ...663..320F} reproduced by permission of the AAS).} 
\label{fig_extobs} 
\end{figure}

The interstellar extinction curves, normalised to the B and V bands as $E(B-V)$ (in the standard manner) as shown in Fig.~\ref{fig_extobs}, seemingly indicate rather wide variations in the extinction curve in the UV region ($\lambda^{-1} > 4 \mu$m$^{-1}$) \cite{2007ApJ...663..320F}. 
However, normalising the extinction curve in this way obviously introduces a strong lever effect because small variations in the extinction at the B and/or V band wavelengths will lead to accentuated UV extinction variations. 
This is perhaps unavoidable when presenting observational data because it is extremely difficult to determine absolute  extinction values. 

As shown by Greenberg \& Chlewicki \cite{1983ApJ...272..563G} normalising the extinction curve in the UV appears to indicate that the FUV reddening is actually rather invariant in shape and intensity. 
They also concluded that the particles responsible for the UV bump at 217\,nm cannot make a significant contribution  to the FUV extinction and that these same particles can only make a small contribution to the extinction long-wards of $\sim 170$\,nm. From this work \cite{1983ApJ...272..563G} it has also been inferred that the FUV extinction carriers ``remain fairly stable once the grains have emerged from the molecular cloud phase of their evolution.'' It was also shown that any dust model for which the FUV extinction is a sum of carbonaceous/graphite and silicate contributions is inconsistent with observations \cite{1983ApJ...272..563G}. 

\begin{figure}[h] 
\begin{centering}
\includegraphics[width=.49\textwidth]{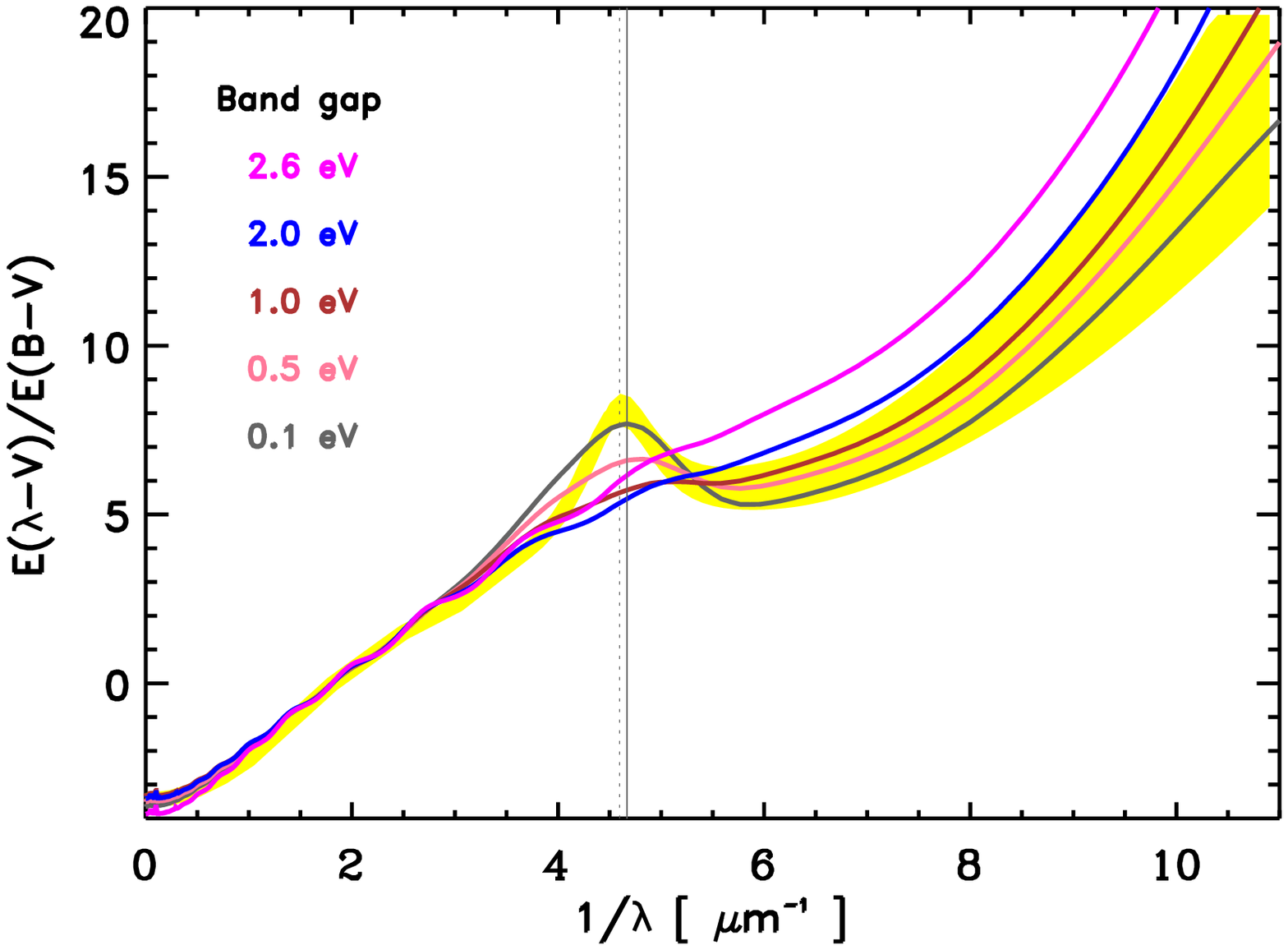} 
\includegraphics[width=.49\textwidth]{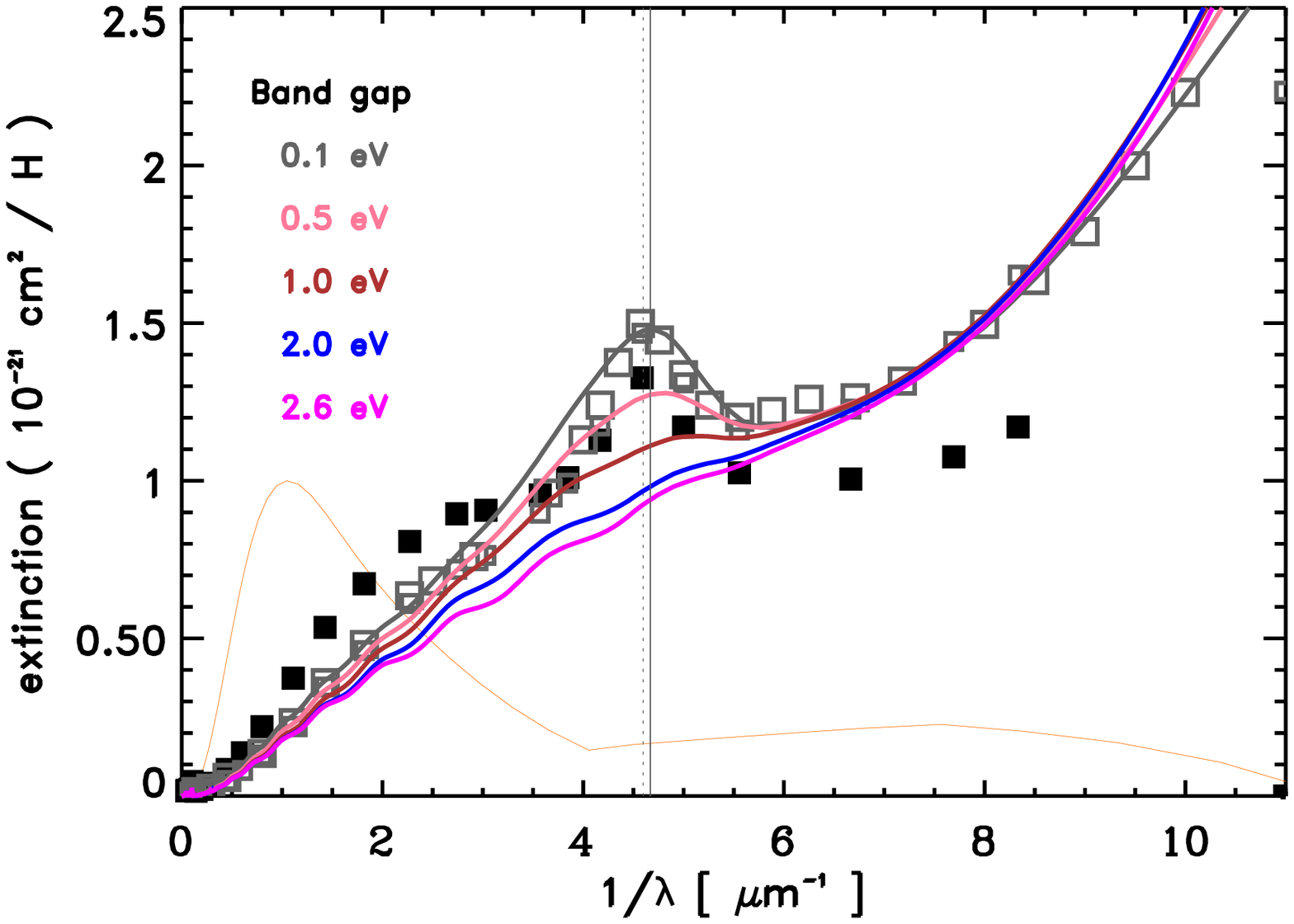}  
\caption{The modelled extinction: $E(B-V)$-normalised (left) and the actual dust cross-sections (right) as a function of the optical properties, which can be characterised by the optical band gap, $E_{\rm g}$ (Figs. 6 and 7 from \cite{2013A&A...558A..62J} reproduced with permission from A\&A). The orange line shows the normalised wavelength dependence of the interstellar radiation field.} 
\label{fig_ext} 
\end{centering}
\end{figure}

Unfortunately, observational data over the entire near-IR to FUV wavelength range is rarely available because of the relative paucity of dust extinction observations in the UV with respect to the visible region. 
However, for a given dust model we can plot the calculated dust cross-sections and hence the absolute or un-normalised extinction. One of the most recent dust models \cite{2013A&A...558A..62J} shows that the FUV extinction is practically invariant for a fixed dust mass and indicates that it is rather the visible extinction and the UV bump that show variations, which depend upon the material composition, and that then skew our interpretation when the data are normalised by $E(B-V)$. Fig.~\ref{fig_ext} shows the standard $E(B-V)$-normalised extinction curve data compared to the dust cross-sections for the Jones {\it et al.} dust model \cite{2013A&A...558A..62J} and clearly illustrates the B and V band bias introduced into normalised data. 

As has been clearly demonstrated \cite{2013A&A...558A..62J}, changes in the carbonaceous dust optical properties, characterised by the material band gap and particle size \cite{2012A&A...540A...1J,2012A&A...540A...2J,2012A&A...542A..98J}, naturally lead to systematic variations in the extinction properties that are more apparent at visible/UV wavelengths than in the FUV-EUV region, unless the abundance of the smallest radius particles ($a \lesssim 3$\,nm) in the dust size distribution is severely perturbed and/or depleted \cite{2013A&A...558A..62J}. An invariance of the dust  properties at short wavelengths (FUV-EUV) reflects that fact that the UV extinction curve simply integrates the summed cross-sections of the atoms incorporated into small ($a < 5 $\,nm) particles \cite{2013A&A...558A..62J}, whatever the chemical composition of the particles. 

Thus, following on from the suggestions of the work by Greenberg \& Chlewicki \cite{1983ApJ...272..563G} it would be more informative to normalise the observed extinction data in the FUV region where the dust cross-sections seemingly show only rather small variations. 

{\it The bottom line}: here is that care should be exercised in the interpretation of interstellar extinction variations based on $E(B-V)$-normalised extinction curves because these can be skewed the assumed normalisation.  

\section{Extinction and the dust size distribution}
\label{sect_dust_ext_sdist}

\begin{figure}[h] 
\begin{centering}
\includegraphics[width=.5\textwidth]{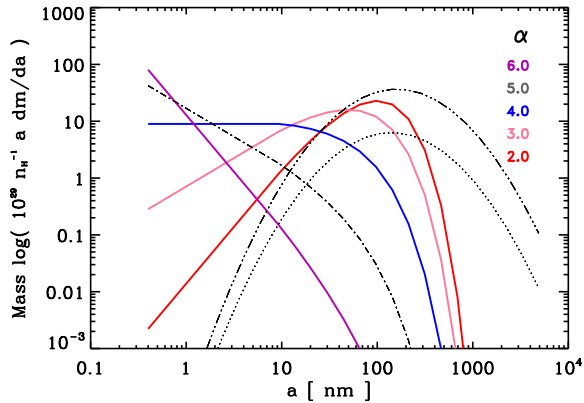} 
\caption{The dust size distributions for the Jones {\it et al.} dust model components: large a-C:H/a-C particles (black dotted), large a-Sil$_{\rm Fe}$/a-C particles (black triple-dot-dashed) and small a-C(:H) carbon particle (black dash-dotted). Variations in the small a-C(:H) carbon particle power-law index, $\alpha$, for fixed dust mass are also shown (coloured lines, Fig. 8 from \cite{2013A&A...558A..62J} reproduced with permission from A\&A).}
\label{fig_sdist} 
\end{centering}
\end{figure}

\begin{figure}[h] 
\begin{center}
\includegraphics[width=.48\textwidth]{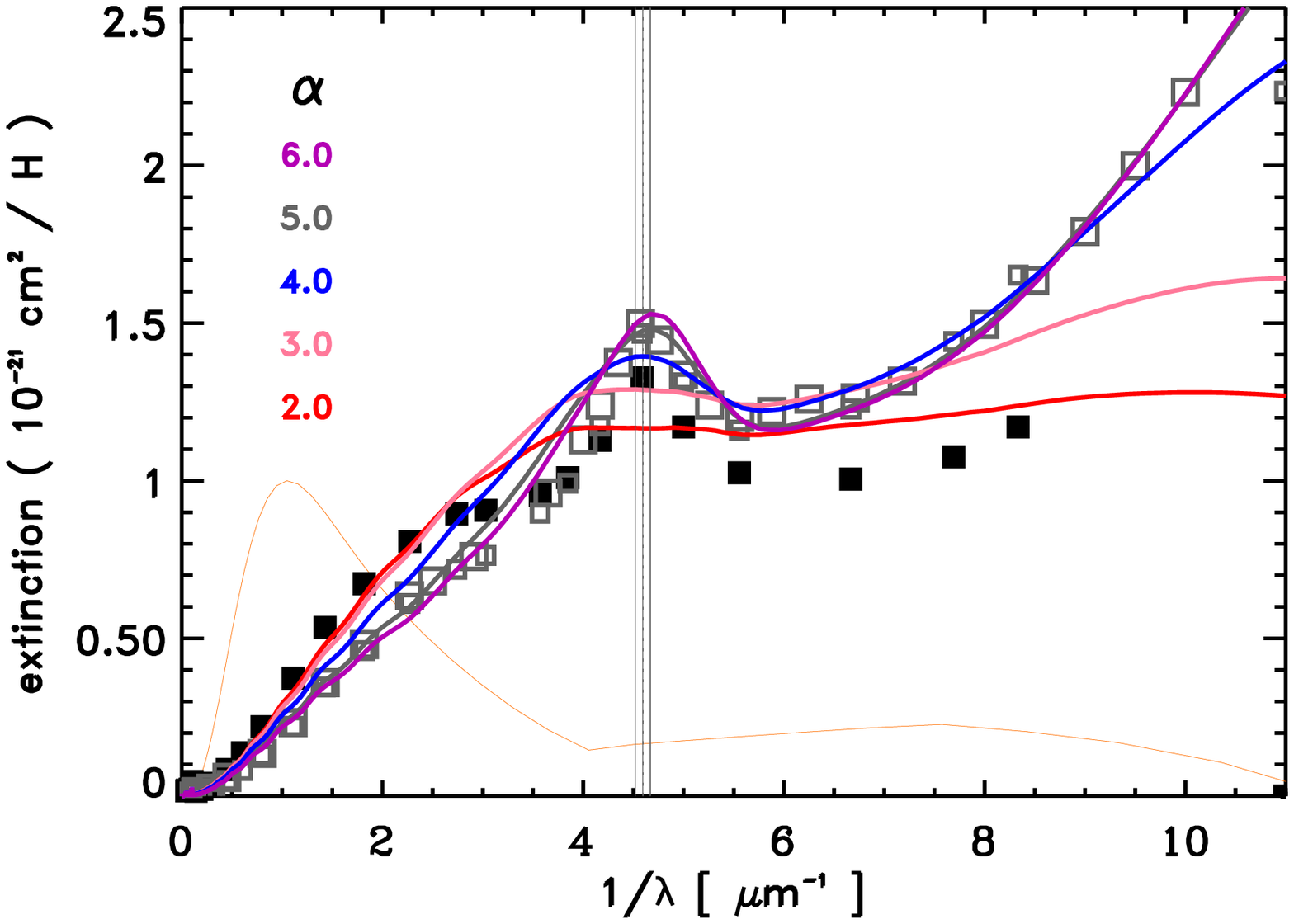} 
\includegraphics[width=.48\textwidth]{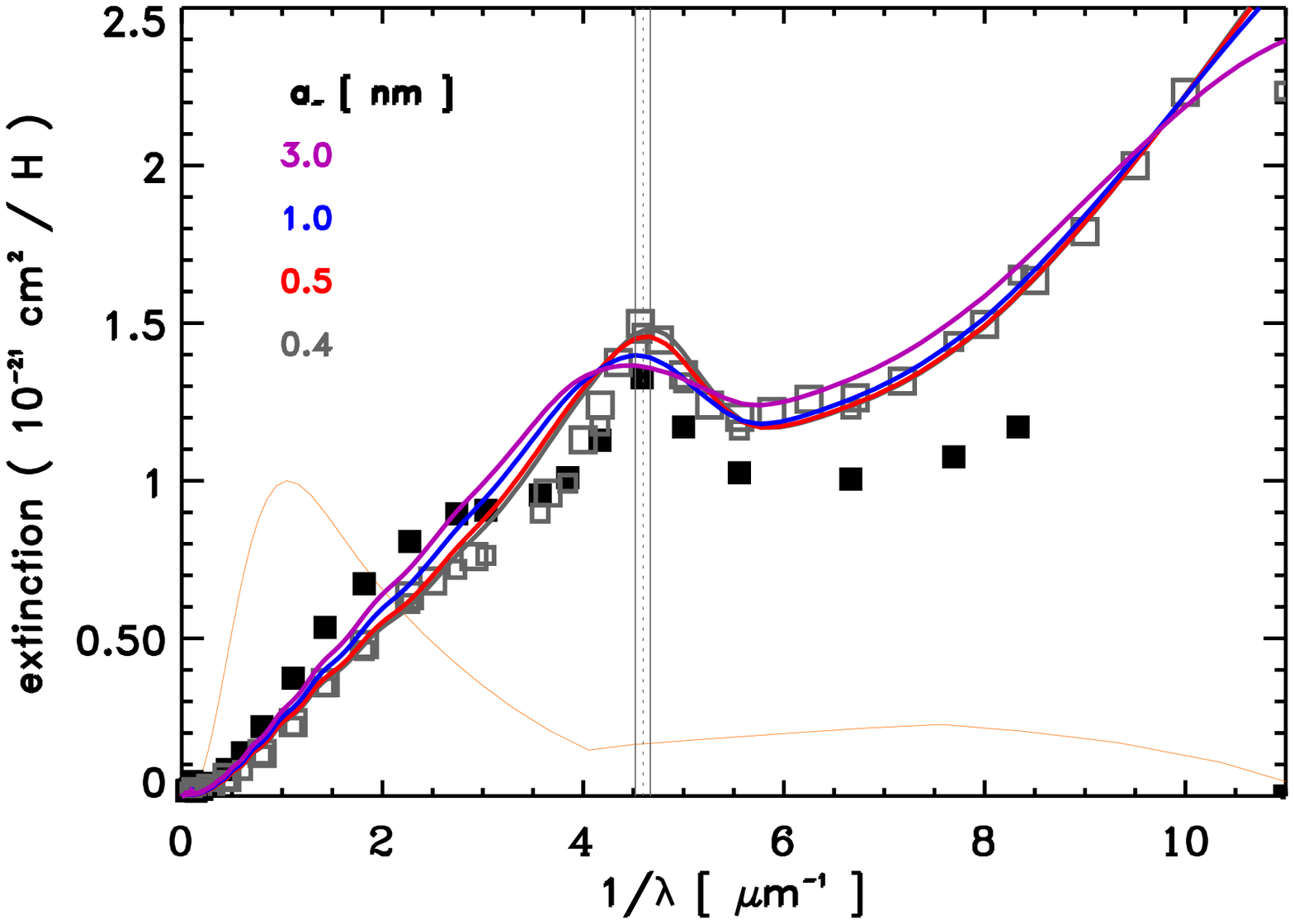} 
\end{center}
\begin{center}
\includegraphics[width=.48\textwidth]{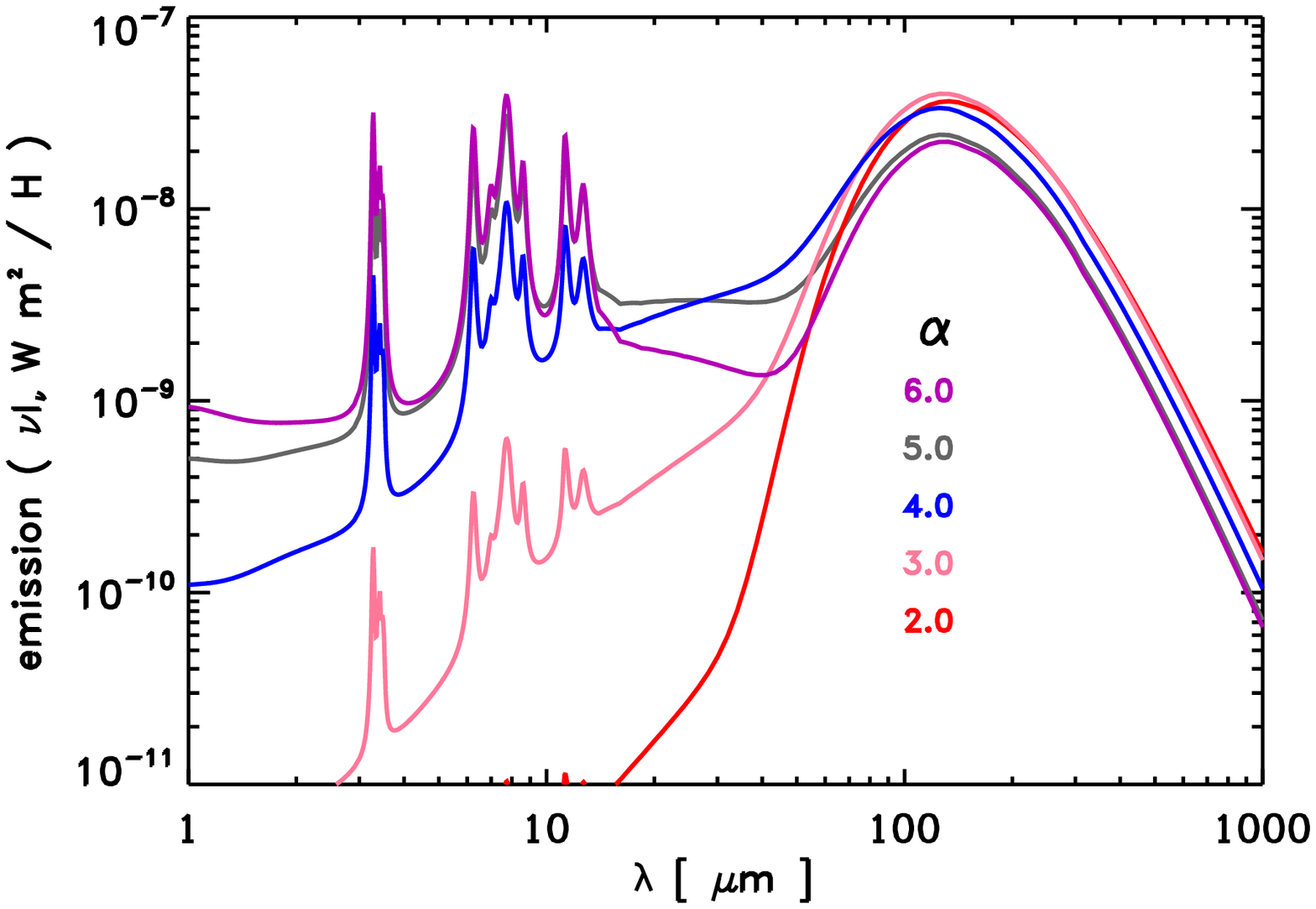} 
\includegraphics[width=.48\textwidth]{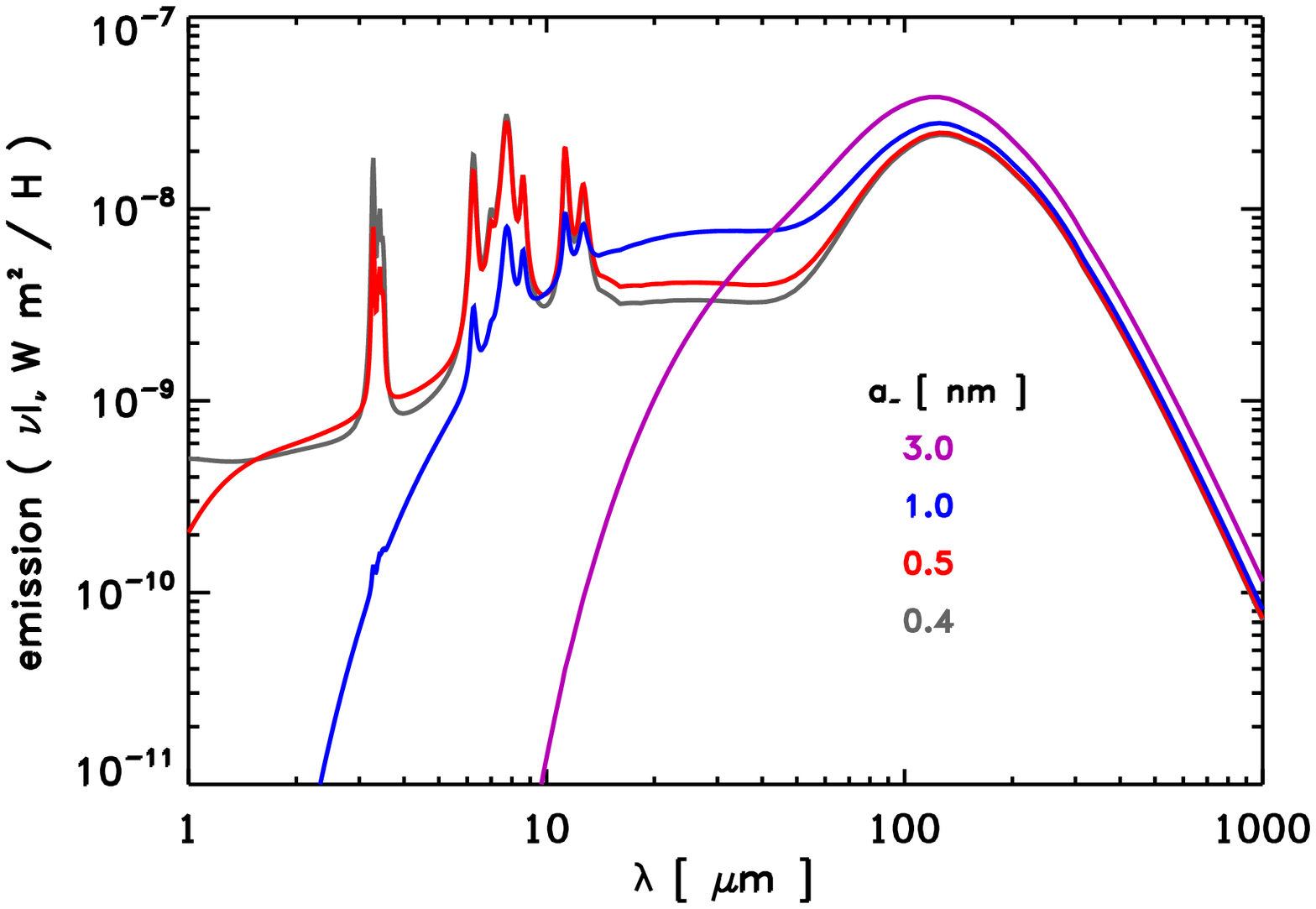}  
\end{center}
\caption{The dust extinction and spectral energy distributions (SEDs) as a function of the the small carbon particle power-law index, $\alpha$, and the minimum particle radius, $a_-$ (Figs. 7 and 12 from \cite{2013A&A...558A..62J} reproduced with permission from A\&A).} 
\label{fig_extem} 
\end{figure}

Here we reflect upon how useful it is to consider extinction as a meaningful and sufficient measure of the dust properties in the ISM. 
In this context we use the recent Jones {\it et al.} dust model \cite{2013A&A...558A..62J} to illustrate how variations in the dust size distribution affect the extinction.  
Fig.~\ref{fig_sdist} shows the adopted dust size distribution for this diffuse ISM model and variants of it with differing small a-C(:H) carbon particle power-law indices, $\alpha$. We will also consider variations due to changing the the minimum assumed particle radius, $a_-$. 
As mentioned in the preceding section, it has been shown that the FUV extinction variations are intrinsically rather limited \cite{1983ApJ...272..563G} and only in the case of a major depletion of small grains from the size distribution (due to the effects of erosion in energetic regions or coagulation in dense regions) should larger variations in the extinction become apparent \cite{2013A&A...558A..62J}.

Fig.~\ref{fig_ext} (right panel) and Fig.~\ref{fig_extem} show the dust cross-sections (upper panels) for a range of values of $\alpha$ and $a_-$ and indicate that, for rather wide variations in the dust optical properties and size distribution, respectively, the extinction curve shows only relatively small variations. These changes clearly do not reflect the model differences because of the degeneracy inherent in extinction-only measurements.
However, if we now consider the spectral energy distribution (SEDs) for the same dust optical properties and size distribution variations (Fig.~\ref{fig_extem}, lower panels) it is evident that they show much greater divergence and therefore seem to better reflect and accentuate dust size distribution variations than the extinction data alone \cite{2013A&A...558A..62J}. It is clear that a study considering only extinction data does not enable us to sufficiently resolve the degeneracies inherent in extinction-alone dust observations and modelling. 
  
Thus, in order to best characterise the observed dust properties the constraints imposed on dust models by, at least,  the observed extinction and emission must be considered and should also take into account the observed linear and circular polarisation ({\it e.g.}, \cite{2014A&A...561A..82S}), albedo and scattering phase function asymmetry ({\it e.g.}, \cite{2004ASPC..309...77G}). 



{\it The bottom line}: is then that studies of extinction alone do not enable us to sufficiently resolve the inherent effect of the degeneracies in the dust size distribution as revealed by extinction modelling alone. Dust extinction studies must therefore be coupled with at least dust emission studies and, ideally, also with complementary information from polarisation, albedo, scattering phase function asymmetry, {\it etc.}

\section{The dust emissivity index ( $\beta$ )}
\label{sect_dust_beta}

The absolute value of the dust emissivity and the slope (wavelength-dependence) of its emissivity at FIR-mm wavelengths, usually indicated as $\beta$, is a direct property of the dust optical properties. $\beta$ is not a free parameter because arbitrarily changing its value, as is often done when modelling dust emission at long wavelengths, implies a change in the dust optical properties and, implicitly, in its composition and structure \cite{2013A&A...552A..89B}.  
The dust emission at FIR to sub-mm wavelengths is often modelled as a modified black-body emission of the form, 
\begin{equation}
I_\lambda = \tau_{\rm \lambda_0} B_\lambda(T_{\rm dust}) \left(\frac{\lambda_0}{\lambda} \right)^\beta, 
\label{eq_emissivity}
\end{equation}
where $\tau_{\rm \lambda_0}$ is the dust opacity at the reference wavelength $\lambda_0$, $B_\lambda(T_{\rm dust})$ is the dust blackbody emission at a temperature $T_{\rm dust}$ and $\beta$ is the dust emissivity index. 
This simple approach has some merit in approximating the measured laboratory variations, which show temperature-dependent spectral slope variations \cite{1996ApJ...462.1026A,1998ApJ...496.1058M,2005ApJ...633..272B}. However, the most recent and most comprehensive measurements on silicates to date, which show clear and significant $\beta-T$ tendencies, also show how the emissivity slopes of Mg-rich amorphous silicates vary with wavelength and do not exhibit simple power-law behaviours \cite{2011A&A...535A.124C}. Further, these authors show that, depending on the form of the Mg-rich silicate, the slope at wavelengths longer than about $500-800\,\mu$m can be flatter (amorphous pyroxene-type) or steeper (amorphous olivine-type) than at the shorter wavelengths \cite{2011A&A...535A.124C,PoSLCDU2013044}. 
Models of low-temperature processes in amorphous silicate materials are able to explain these laboratory-measured variations  \cite{2007A&A...468..171M}. However, further work is yet needed in order to ascertain if this same range of emissivity behaviours also translates to Fe-rich silicates. Thus, it appears that we still have much to learn about the low-temperature physics relating to interstellar amorphous silicate grain analogues. We should therefore remain rather cautious in our interpretation of the emission from the large cold grains in the ISM until such time as we have a better handle on their composition (olivine-type, pyroxene-type or other mineralogies) and structure (homogeneity, porosity or mantling). 

Recent observations of dust in the diffuse ISM with the {\em Planck} mission \cite{2013arXiv1312.1300P} indicate a mean dust temperature of $19.7 \pm 1.4$\,K and dust emissivity index $\beta = 1.62 \pm 0.10$. 
In comparison, laboratory measurements \cite{2011A&A...535A.124C} and dust modelling \cite{2013A&A...558A..62J} show that single-valued dust emissivity slopes, {\it i.e.}, $\beta$ constant, are unlikely and need to be replaced by some form of smoothly varying function across the FIR to mm wavelength regime. 
Thus, $\beta$ is not a single-valued, free parameter; changing it changes $\kappa(i,\lambda)$ (the mass absorption coefficient of the material $i$) because it depends upon the composition and structure of the dust material, its temperature and the wavelength under consideration, {\it i.e.}, in reality we have $\beta(i,T_{\rm dust},\lambda, \ldots)$. So, in retrospect it seems to have been a rather na\"ive assumption that $\beta$ would be constant and independent of material, temperature and wavelength. 

{\it The bottom line}: is that single-$\beta$ modified blackbodies are not a good approximation to the dust emission in the ISM and that any derived $T_{\rm dust}$ and  $\beta$ values, or range of values, are unlikely to be physically-meaningful in terms of real dust temperatures. They are also going to be something of a compromised fit to real dust SED observational data.  Also, our current knowledge of the `real' nature of the large amorphous silicate grains in the ISM is far from complete but should perhaps be better guided by the analysis of the pre-solar silicate grains extracted from primitive meteorites \cite{PoSLCDU2013040}. 


\section{The elemental composition of dust}
\label{sect_dust_comp}

\begin{figure}[h] 
\begin{center}
\includegraphics[width=.5\textwidth,angle =270.]{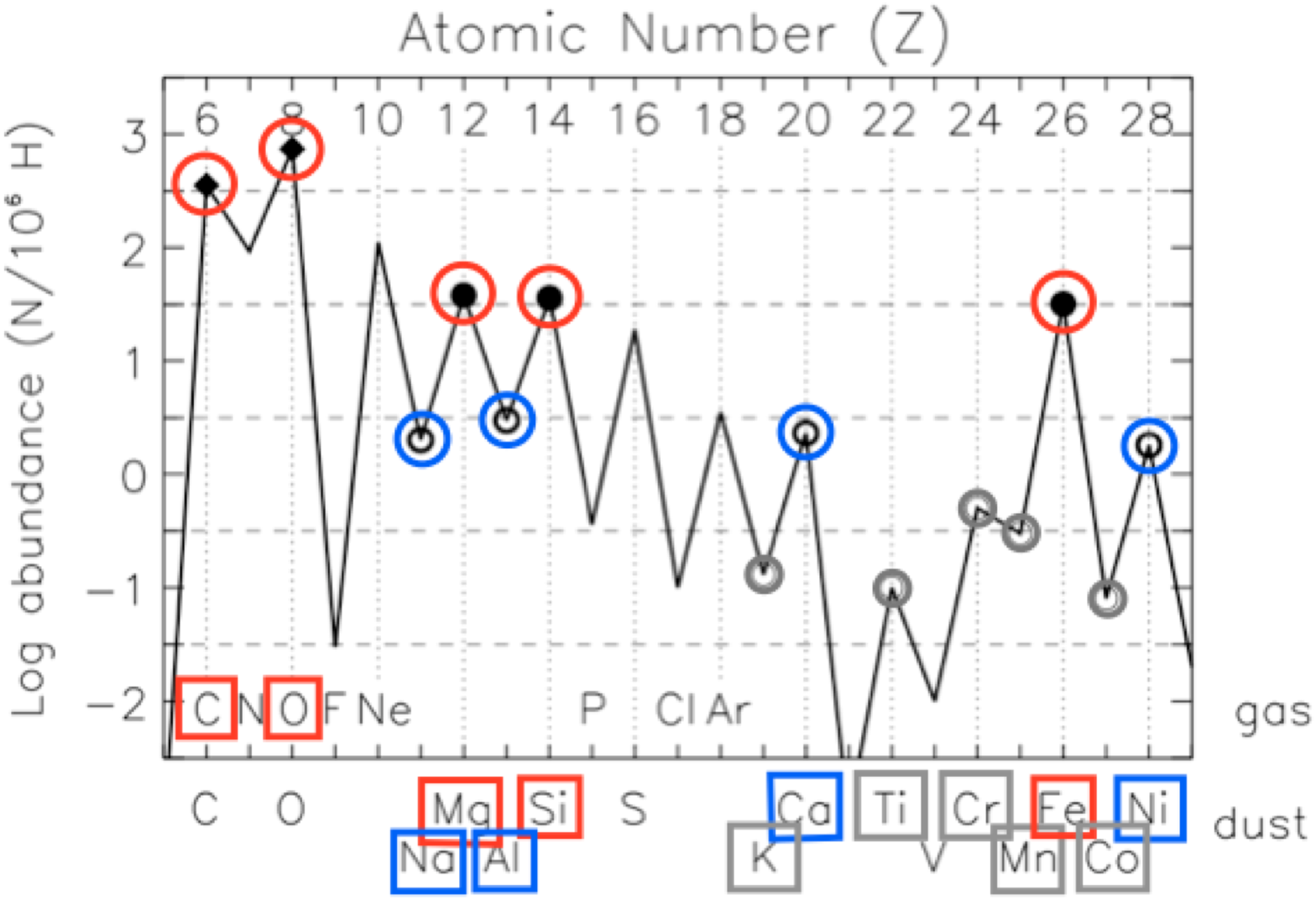} 
\caption{The relative abundances of the elements (in ppm with respect to H) as a function of atomic number. The most abundant silicate/oxide and solid (hydro-)carbon dust-forming elements (red), and the minor (blue) and trace (grey) dust-forming elements are highlighted.} 
\label{fig_depletions} 
\end{center}
\end{figure}

Fig.~\ref{fig_depletions} summarises the abundances relative to hydrogen of the most important interstellar dust-forming elements along with other, more volatile elements. 
Depletion studies \cite{2009ApJ...700.1299J,PoSLCDU2013005}, 
indicate that the dust elemental composition is dominated by C, O, Si, Mg and Fe with minor amounts of Na, Al, Ca and Ni and traces of K, Ti, Cr, Mn, and Co. In addition, chemical considerations indicate that these elements are probably bound into two major dust phases, one C-rich and the other O-rich. The O-rich phase most likely consists of (amorphous) silicates and/or oxides predominantly composed of O, Si, Mg and Fe, while the C-rich phase is probably mostly of carbon but must also contains some H atoms, most likely bound to C atoms in some form of hydrogenated amorphous carbon solid (rather than graphite). 
While interstellar elemental depletion studies are able to indicate the stoichiometric composition of the condensed solid phase in the ISM they reveal nothing of the dust mineralogy. One rather intriguing conundrum relating to the amorphous silicate/oxide phase is that there is to date little, if any, clear IR spectroscopic evidence for the presence of Fe in dust in the ISM, around evolved stars or in circumstellar discs.  So, where exactly is the Fe hiding and in exactly what solid form is it bound? The obvious Fe-incorporating candidate solids are silicate, oxide, sulphide or metal particles. In most cases, where amorphous silicates have been crystallised (around evolved stars or in circumstellar discs), their composition appears to be Mg-rich and there is little direct evidence for Fe oxides, which leaves sulphides and metal as the best candidates. A mix of Fe metal and sulphide could temptingly be identified with the thorny problem of the origin of the GEMS (Glass with Embedded Metal and Sulphides) but this component of the interplanetary dust particles (IDPs) has now been demonstrated to be of solar nebula origin \cite{PoSLCDU2013040}. 
The problem of interstellar Fe is discussed again within the framework of interstellar silicates in \S\,\ref{sect_silicate_dust}.


{\it The bottom line}: is that the absolute elemental abundances and the depletion of elements into dust in the ISM gives us a measure of the dust stoichiometry but not its exact chemistry, structure or composition

\section{The cosmic carbon abundance}
\label{sect_cosmic_C}

\begin{figure}[h] 
\begin{center}
\includegraphics[width=.5\textwidth,angle =270.]{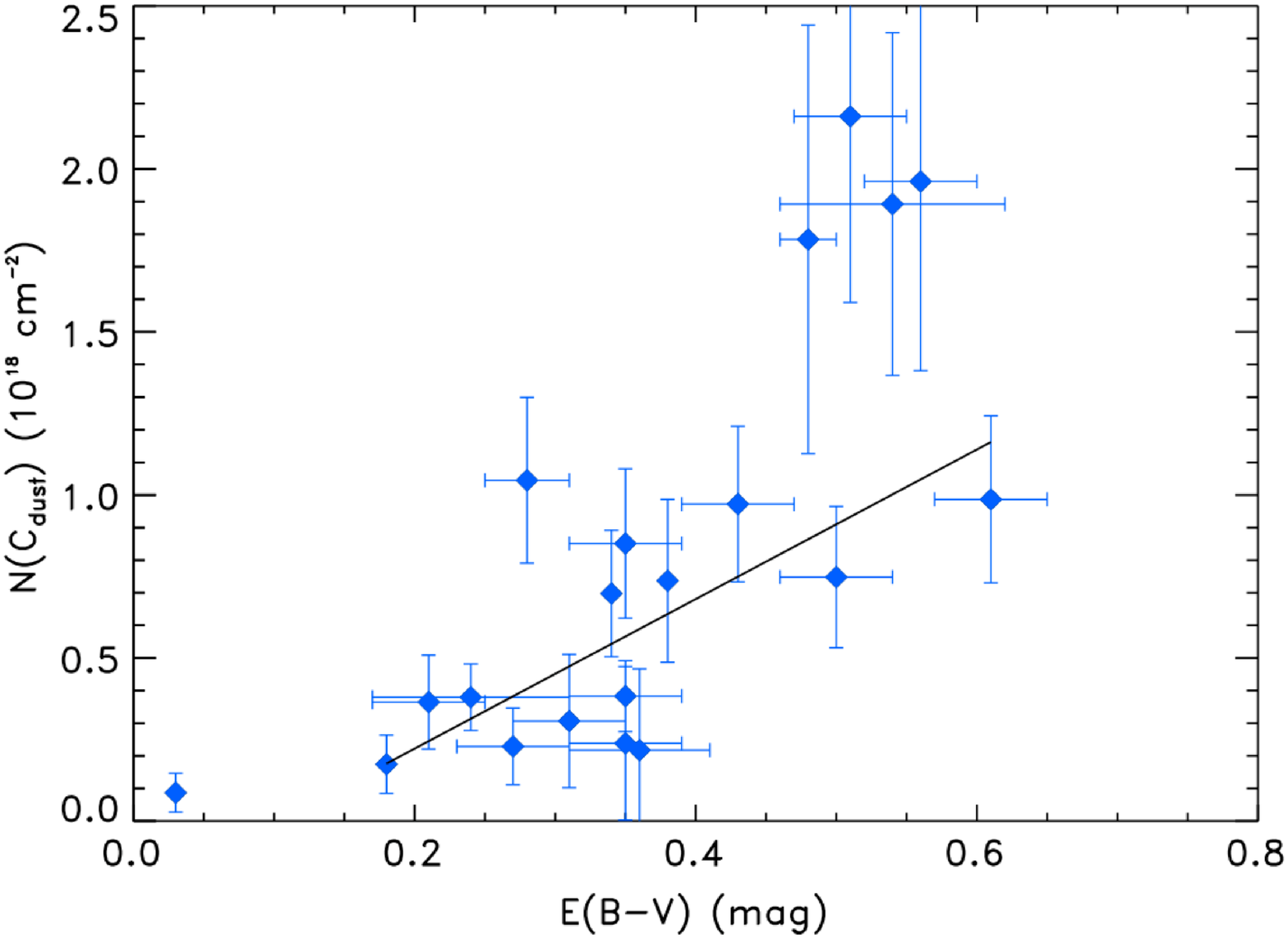} 
\caption{The column density of C in dust, N(C$_{\rm dust}$), as a function of $E(B-V)$, with the uncertainties on N(C$_{\rm dust}$) and $E(B-V)$ indicated. The linear least-squares fit to the correlation between N(C$_{\rm dust}$) and $E(B-V)$ excludes the lower left data point and has a correlation coefficient of 0.71 (Fig. 4 from \cite{2012ApJ...760...36P} reproduced by permission of the AAS).} 
\label{fig_Cdustcol} 
\end{center}
\end{figure}

The latest measurements of the carbon abundance in the ISM estimate its total abundance to be of the order of at least $470$\,ppm \cite{2012ApJ...760...36P}, which is considerably higher than previous measures and thus eases this key abundance constraint dust models. 
In fact it seems that practically all of the available dust models are consistent with these new data in low $n_{\rm H}$  regions but that they do not use enough carbon in dust in the high $n_{\rm H}$ regions of the ISM \cite{2012ApJ...760...36P}.  
As these same authors show (see Fig.~\ref{fig_Cdustcol}), the column density of C in dust, $N({\rm Cdust})$, increases steadily up to $N({\rm Cdust}) \simeq 10^{18}$\,cm$^{-2}$ with increasing extinction up to $E(B-V) \simeq 0.4$. Thereafter, for $0.4 \lesssim E(B-V) \lesssim 0.6$, the data show a spread in $N({\rm Cdust})$ by about a factor of two ($N({\rm Cdust}) \simeq 1-2 \times 10^{18}$\,cm$^{-2}$). This perhaps indicates that C incorporates into dust in the denser regions into a more volatile carbonaceous dust species and that this carbonaceous dust is different from the typical carbon-rich dust in the more diffuse ISM (see \S\,\ref{sect_dust_spect} for a more detailed discussion of this issue). 

{\it The bottom line}: is that there is no and never was a carbon crisis, {\it i.e.}, carbon now seems to be so abundant in the ISM that all current dust models can be accommodated. However, in the denser regions of the ISM it appears that none of the current models use enough carbon in dust. 

\section{Dust spectroscopic features}
\label{sect_dust_spect}

The chemical composition and mineralogical properties of dust can be constrained by characteristic spectroscopic bands observed in absorption and emission throughout the ISM, which indicate that it principally consists of amorphous silicates and aromatic-rich and/or aliphatic-rich amorphous carbonaceous matter. One of the principal methods in the analysis of these spectral bands has been to subtract some form of underlying continuum in order to extract the details of the particular band shapes. However, independent of whether the bands are in absorption or emission, the removal of an underlying `continuum' is less than ideal because essential information is lost, especially if some or all of the underlying `continuum' is actually due to emission from the band carriers themselves. In order to illustrate the difficulty a particular example of interstellar absorption spectroscopy will now be considered; the interstellar aliphatic $\sim 3.4\,\mu$m absorption band extending over the $3.2-3.6\,\mu$m range. 

\begin{figure}[h] 
\includegraphics[angle=90.,  width=.5\textwidth]{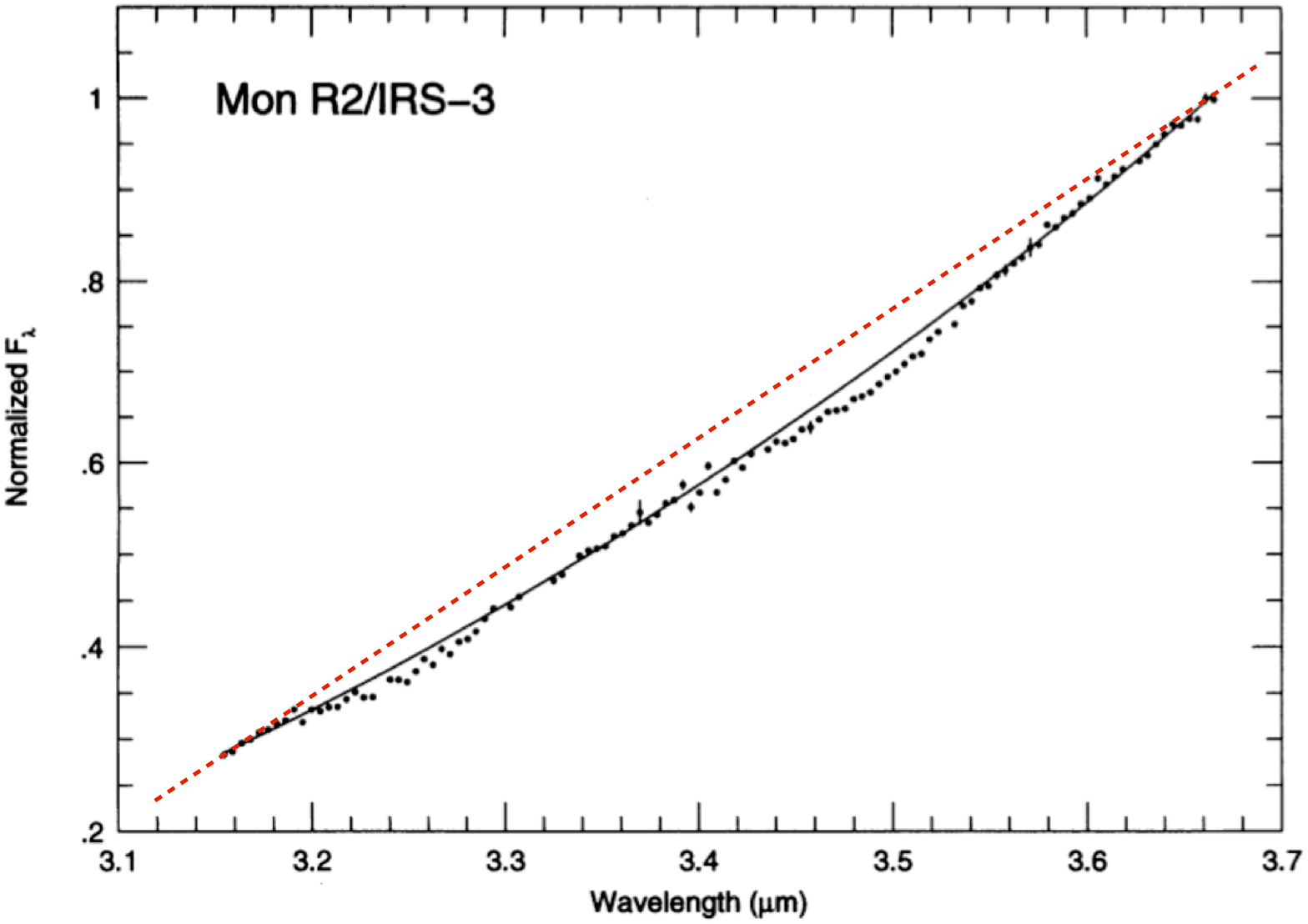} 
\includegraphics[                   width=.48\textwidth]{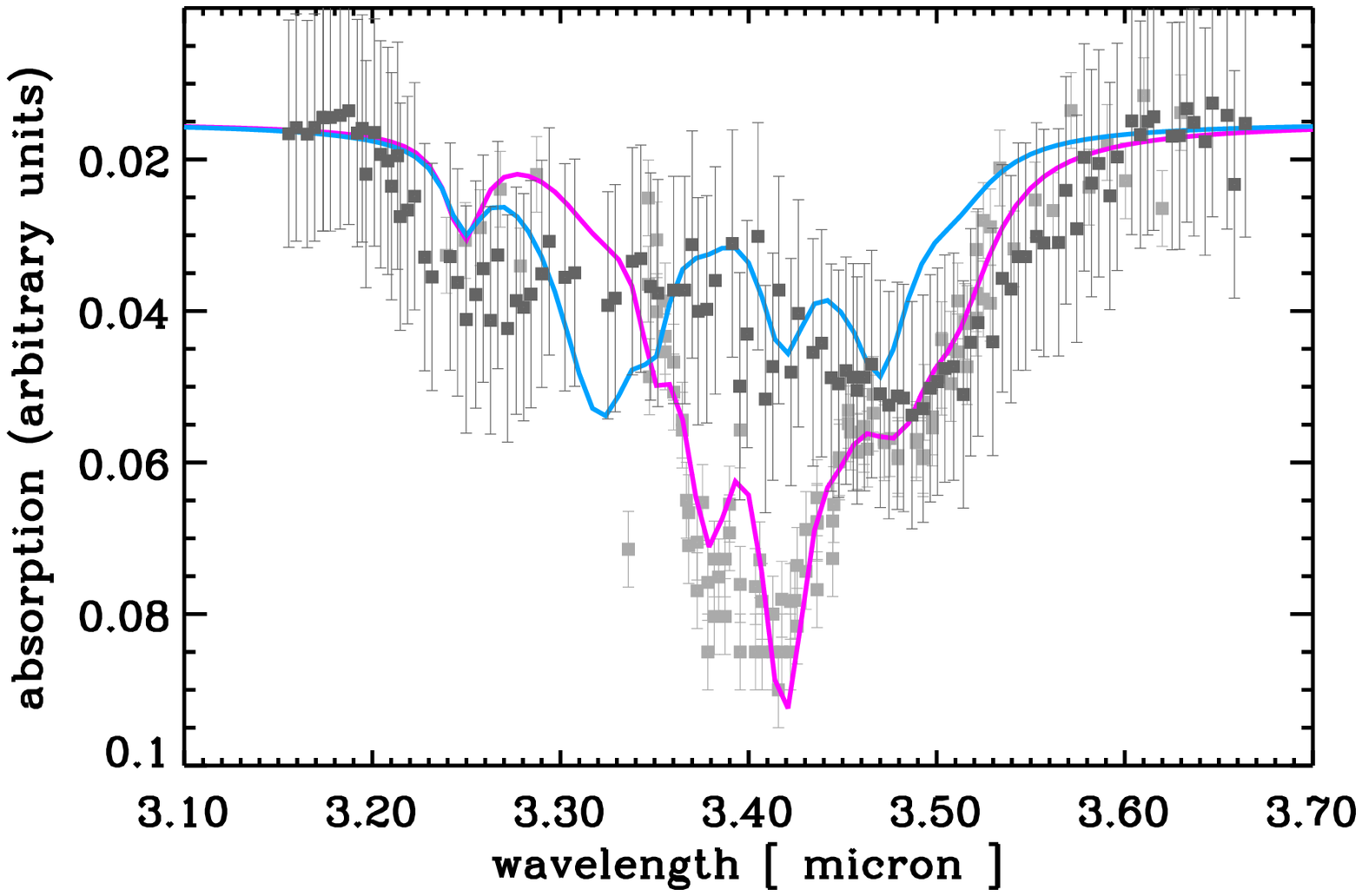} 
\includegraphics[angle=270.,width=.5\textwidth]{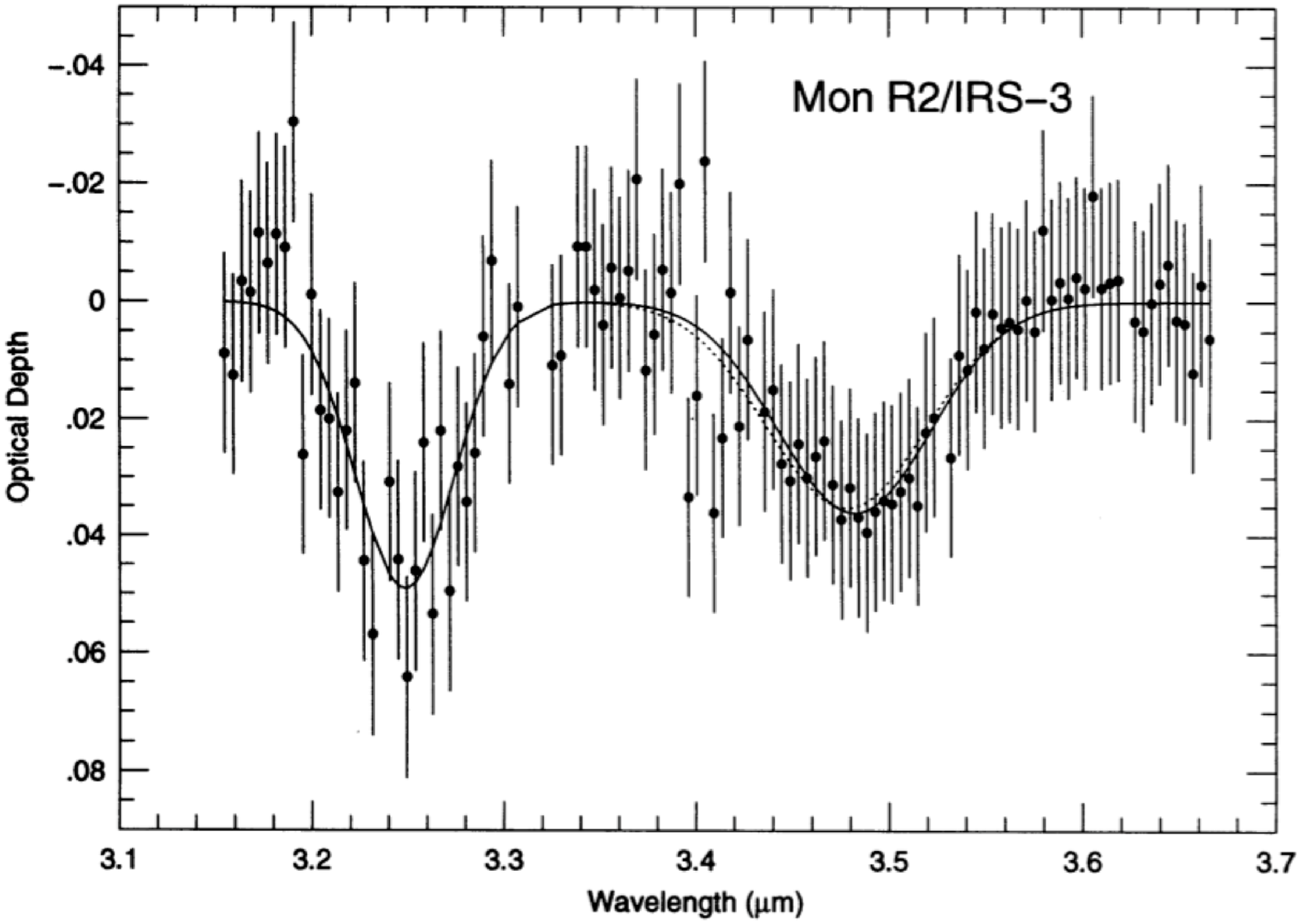} 
\caption{Top left: The $3.1-3.7\,\mu$m portion of the long wavelength wing of the $3\,\mu$m ice absorption band towards the protostar Mon R2/IRS3 (Fig. 1 from \cite{1994ApJ...433..179S} reproduced by permission of the AAS) with the assumed underlying continuum (black line). Also shown is an alternative linear baseline (dotted red line). 
Bottom: The continuum-subtracted $3.1-3.7\,\mu$m spectrum (Fig. 2 from \cite{1994ApJ...433..179S} reproduced with the permission of the AAS). 
Top right: The same spectrum with the linear baseline shown in the top left panel subtracted from the data (dark grey squares). For comparison the light grey squares show the diffuse ISM spectrum towards the Galactic Centre sources IRS6E and Cyg OB2 No. 12 \cite{2002ApJS..138...75P}. The blue (violet) line shows the absorption spectrum of an aliphatic-rich a-C:H with $E_{\rm g} = 2.25$\,eV ($2.5 $\,eV).}
\label{fig_3mic_abs} 
\end{figure}

The $\sim 3.4\,\mu$m absorption band is actually composed of many sub-bands, of which the most evident in the diffuse ISM occur at $3.38$ and $3.42\,\mu$m and are attributed to aliphatic CH$_3$ and CH$_2$ asymmetric C$-$H stretching modes, respectively (see the light grey data points in the top right panel of Fig~\ref{fig_3mic_abs}).
In comparison, an analysis of the observed interstellar $\sim 3.4\,\mu$m absorption in the dense, molecular ISM towards protostars appears to lack the sub-bands at $\simeq 3.38$ and $\simeq 3.42\,\mu$m and instead exhibit two clear bands at $\simeq 3.25$ and $\simeq 3.48\,\mu$m with little absorption in between (Fig~\ref{fig_3mic_abs}, bottom panel, Fig. 2 from \cite{1994ApJ...433..179S}). This result seems to implies that the carbonaceous material along these lines of sight is very different from that in the diffuse ISM \cite{1994ApJ...433..179S}.

However, a close look at the curved baseline subtracted from the dust continuum towards the protostar Mon R2/IRS3 (Fig~\ref{fig_3mic_abs}, top left panel: Fig. 1 from \cite{1994ApJ...433..179S}) shows that it closely follows the shape of the absorption band, as compared to an alternative linear baseline. Given that the $\sim 3.4\,\mu$m absorption feature actually lies on the red wing of the deep and much stronger $\sim 3.1\,\mu$m ice absorption band \cite{1992ApJ...399..134A}, the adoption of a concave, rather than a linear or even convex baseline, seems to remove key information from the original spectroscopic data. The removal of a linear baseline (the dotted red line in Fig~\ref{fig_3mic_abs}, top left panel) from the data leads to the $\sim 3.4\,\mu$m absorption band profile, shown by the dark grey data points in the top right panel of Fig~\ref{fig_3mic_abs}, that is rather different from the band profile originally shown \cite{1994ApJ...433..179S} and appears to be rather plateau-like between the $3.25$ and $3.48\,\mu$m sub-peaks rather than falling to zero in the intervening region. Thus, the observed $\sim 3.4\,\mu$m band profile in the dense, molecular regions is indeed clearly different from that in the diffuse ISM. However, the long-held view that the $\simeq 3.38$ and $\simeq 3.42\,\mu$m features are absent in dense regions appears to be untenable. A better interpretation of the data would be that these bands are still present in dense regions but that they are much weaker and lead to a more plateau-like $\sim 3.4\,\mu$m absorption feature, which appears to consistent with slightly less aliphatic-rich solid hydrocarbons with a band gap close to 2\,eV (see Fig~\ref{fig_3mic_abs}, top right panel) but is much more aliphatic than the bulk of the carbonaceous dust in the diffuse ISM with a band gap close to 0\,eV \cite{2013A&A...558A..62J,2012A&A...540A...1J,2012A&A...540A...2J,2012A&A...542A..98J}. 

A related issue is that the observational data on interstellar amorphous silicates, {\it e.g.}, the $\simeq 9.7$ $\sim 18\,\mu$m absorption bands, are very often presented in the form of continuum-subtracted spectra, which are then analysed in order to derive silicate compositional/structural data. However, given that it is very hard to define the limits of these broad bands, it is likely that any continuum-subtracted spectrum will not be the best approximation of the real band profiles. 

Thus, and in conclusion, in the interpretation of IR spectra it is always preferable to undertake full modelling studies because the observed bands and underlying continua will almost always have a common origin, {\it i.e.}, the two are physically related and must therefore be treated together in a full analysis. 

{\it The bottom line}: is that extreme care must be exercised in the interpretation of the data yielded by baseline-subtracted spectra because the adopted form of the baseline or underlying continuum can lead to a removal of key spectral information and hence to a misinterpretation of the data. Thus, full dust modelling is always to be preferred. 

Further, a re-evaluation of the $\sim 3.4\,\mu$m absorption band profile through the molecular ISM towards the protostar Mon R2/IRS3 indicates that the hydro-carbon dust  in this dense region is clearly different from that in the diffuse ISM but that it is consistent with the evolution of the IR spectral properties of a-C:H materials.

\section{The nature of interstellar carbonaceous dust}
\label{sect_C_dust}

For more than 35 years graphite been used as the carbon material of choice in the most widely-used interstellar dust models \cite{1977ApJ...217..425M,2014A&A...561A..82S,1984ApJ...285...89D,2001ApJ...551..807D,2001ApJ...554..778L,2007ApJ...657..810D}; with a few exceptions \cite{2013A&A...558A..62J,1997A&A...323..566L,2004ApJS..152..211Z,2011A&A...525A.103C}. However, graphite has never been detected in the ISM and it is not one of the most abundant pre-solar grain species (see the paper by Scott Messenger in these proceedings). Thus, its usefulness as a carbon dust analogue is questionable and it is perhaps now time to adopt something more physical and all the complex baggage that that entails. 
In this respect, amorphous hydrocarbon solids, a-C(:H)
are intriguing and present a challenge because of their surprising complexity \cite{2012A&A...540A...1J,1986AdPhy..35..317R,1987PhRvB..35.2946R,1988PMagL..57..143R,2000PhRvB..6114095F,2004PhilTransRSocLondA..362.2477F}. 
In particular, they are susceptible to UV photo-processing and thermal annealing \cite{2012A&A...540A...1J,2012A&A...540A...2J,2012A&A...542A..98J,2009ASPC..414..473J,2011A&A...530A..44J}. 
Further, incident ion and electron collisions in shock waves, due to cosmic rays and in a hot gas can lead to the rapid destruction of a-C(:H) nano-particles \cite{2008A&A...492..127S,2010A&A...510A..36M,2010A&A...510A..37M,2011A&A...526A..52M,2012A&A...545A.124B}. 
Amorphous (hydro)carbon materials have already been considered as the primary carbon dust phase in a number of ISM dust models \cite{2013A&A...558A..62J,1998ApJ...496.1058M,2004ApJS..152..211Z,2011A&A...525A.103C,1995ApJS..100..149M,1996ApJ...464L.191M,2001A&A...367..355M}. 

Hydrogenated amorphous carbon materials cover a wide range of compositions, from wide band gap, H-rich, aliphatic-rich a-C:H to narrow band gap, H-poor, aromatic-rich a-C. 
The optical properties of this suite of materials can, principally, be characterised by a single characteristic, their band gap ($E_{\rm g} = -0.1$ to 2.7\,eV), which is directly proportional to $X_{\rm H}$, the H atom fraction, {\it i.e.}, $E_{\rm g} \simeq \,4.3 X_{\rm H}$ \cite{1990JAP....67.1007T}, where $X_{\rm H} = X_{\rm H}/(X_{\rm C}+X_{\rm H})$ and $X_i$ is the element $i$ atomic fraction.  
Another key characteristic, which is a function of $E_{\rm g}$ and $X_{\rm H}$, is the ratio, $R$, of the $sp^3$ and $sp^2$ C atomic fractions, $X_{sp^3}$ and $X_{sp^2}$, respectively. 
In random covalent network (RCN) models \cite{1979PhRvL..42.1151P,1980JNS...42...87D,1983JNCS...57..355T,1988JVST....6.1778A} and extended RCN (eRCN) models \cite{2012A&A...540A...1J,1990MNRAS.247..305J} $X_{\rm H}$, $E_{\rm g}$ and $R$ are related by the expressions,  
\begin{equation}
R = \frac{X_{sp^3}}{X_{sp^2}} \approx \frac{(8\,X_{\rm H}-3)}{(8-13\,X_{\rm H})} \sim \frac{(0.6\,E_{\rm g}-1.0)}{(2.7-E_{\rm g})};  
\label{eq_REg}
\end{equation}
which is valid for a-C:H with $0.4 \lesssim X_{\rm H} \lesssim 0.6$ or equivalently $1.5$\,eV $\lesssim E_{\rm g} \lesssim 2.7$\,eV. For more highly-structured, aromatic-rich a-C materials the RCN models are not appropriate and other approaches based on defective graphite networks \cite{1990JAP....67.1007T} or full size-dependent surface/network models are required \cite{2012A&A...540A...1J}. 
Note that Eq.\,(\ref{eq_REg}) is only valid for a-C:H and is an approximation to the exact relationship between $E_{\rm g}$, $X_{\rm H}$ and $R$, which depend upon the $sp^2$ aromatic domain sizes, the -CH$_3$ methyl group concentration and the particle size \cite{2012A&A...540A...1J,2012A&A...540A...2J,2012A&A...542A..98J}. 

The recent Jones {\it et al.} model \cite{2013A&A...558A..62J} is an attempt to harness the laboratory-constrained thermal and optical properties of a-C(:H) solids, and the inherent variations in their optical properties, into a new view of dust in the ISM. The result is a model that is coherent with many dust observables, their variations and inter-correlations and, perhaps most significantly, predicts that dust must evolve in response to the local physical conditions, principally through the effects of UV photo-processing (see Figs. \ref{fig_ext}, \ref{fig_extem} and \ref{fig_3mic_abs} \cite{2013A&A...558A..62J,2012A&A...542A..98J}). 


{\it The bottom line}: is that graphite grains are not an important component of interstellar dust.  We must instead consider the interstellar carbonaceous dust to be primarily composed of the suite of materials collectively known as hydrogenated amorphous carbons, a-C(:H) or HAC.  

\section{Dust luminescence}
\label{sect_luminescence}

In addition to the dust extinction, emission and polarisation observed in the ISM and in circumstellar regions some component of the dust luminesces in the red (extended red emission, ERE) and in rare cases also in the blue (blue luminescence, BL). 
The Red Rectangle is an interesting proto-planetary nebula that displays both ERE and BL. 
In its central regions an accretion disc around a main sequence secondary star is fed by overflow from the post-AGB primary, giving rise to an internal Lyman/far-UV continuum \cite{2009ApJ...693.1946W}. 
This inner system is surrounded by an optically thick, edge-on disc, which attenuates forward-scattered radiation \cite{2006ApJ...653.1336V}.  
In the Red Rectangle the ERE is observed to peak close to the central star (HD\,44179) in the outflow cavity and walls, which are exposed to the FUV photons from the central regions \cite{2006ApJ...653.1336V}. 
Conversley, the BL is extended and associated with the outermost parts of the disc that are shielded from the internal FUV photons \cite{2004ApJ...606L..65V,2005ApJ...633..262V,2005IAUS..235P.234V} but are exposed to UV photons from the ambient interstellar radiation field. 

It has been noted that the sharp emission lines superimposed on the ERE in the Red Rectangle region resemble the zero phonon lines of terrestrial diamonds \cite{1989ASS...150..387D}, indicating that $sp^3$-rich, wide band gap carbonaceous dust is an important dust component here.
It is also clear that the ERE is not strongly associated with the IR emission bands (usually associated with aromatic-rich carbonaceous materials) but it does correlate with the FUV ($E > 10.5$\,eV), which is required for its excitation \cite{1985ApJ...294..225W,2006ApJ...636..303W}. 
It is noticeable that FUV photons are also required to de-hydrogenate and aromatise $sp^3$-rich carbonaceous solids, {\it i.e.}, a-C:H \cite{2012A&A...540A...1J,2012A&A...540A...2J,2012A&A...542A..98J}. 

The unusual configuration of the Red Rectangle region could help to explain the mutual exclusivity of the inner ERE and outer BL spatial distributions.
In the inner regions the FUV continuum photons transform $sp^3$-rich a-C:H dust into $sp^2$-rich a-C dust and re-configure it into ERE carriers, which are excited by UV photons. Further out $sp^3$-rich a-C:H dust in the outer regions of the disc is not transformed and it is this aliphatic-rich material that is responsible for the higher energy BL, which is probably excited by the interstellar UV photons.
After injection into the ambient ISM the $sp^3$-rich a-C:H to a-C transformation will occur over timescales of the order of $10^5-10^6$\,yr  \cite{2012A&A...542A..98J} and so the BL is not expected to extend far beyond the bounds of the Red Rectangle region. 

In the laboratory it has been seen that some thin films of tetrahedral $sp^3$-rich, diamond-like a-C:H, with and without incorporated nitrogen atoms (taC:H and ta-C:N), exhibit a broad BL with superimposed bands \cite{2006TSF...515.1597P}.  It is interesting that one of the superimposed bands in the taC:N blue luminescence occurs at 442.8\,nm, the same wavelength as the strongest diffuse interstellar band (DIB). 
Thus, given the Red Rectangle observations, the photo-processing of N-doped a-C:H materials could provide a route DIB carrier candidates \cite{2013A&A...555A..39J,Jones_CosmicDust_PSS}. 
A photo-processing scheme was indeed proposed for the formation of the 580\,nm emission line carrier in the Red Rectangle, which could be related to the 579.7\,nm DIB \cite{1998MNRAS.301..955D}. 
It is also of note that the carrier of the 579.7\,nm DIB towards $\chi$\,Vel  is more sensitive to the interstellar UV radiation field than to the local density \cite{2013MNRAS.429..939S}.
Nevertheless, and despite the strong evidence for a-C:H photo-processing in the Red Rectangle, only two weak DIBs (578.0 and 661.3\,nm) have been observed, originating either in the Red Rectangle itself or in the intervening diffuse ISM \cite{2004ApJ...615..947H}. 

{\it The bottom line}: is that the properties of $sp^3$-rich (hydro-)carbon dust do seem to evolve in a systematic way in response to  FUV photon irradiation and that the carbonaceous material evolutionary sequence is from H-rich a-C(:H) to H-poor a-C as a result of photo-processing in the ISM and circumstellar regions.

\section{The interstellar `silicate' composition}
\label{sect_silicate_dust}

The idea that a significant fraction of the interstellar dust mass exists in the form of a special kind of ``astronomical'' silicate has served us well for about 30 years \cite{1984ApJ...285...89D}. However, it should be remembered that this material was constructed from and to fit observations \cite{1984ApJ...285...89D} because the appropriate laboratory data were not available at that time. Happily we now have abundant data on laboratory analogues of likely interstellar silicate materials 
\cite{1996ApJ...462.1026A,1998ApJ...496.1058M,2005ApJ...633..272B,2011A&A...535A.124C}, which should enable us to better constrain the likely properties of any interstellar silicate material. Unfortunately, the interpretation of the measured variations in the absolute value and wavelength-dependence of the emissivity of these silicate analogues is not straightforward \cite{2007A&A...468..171M,PoSLCDU2013044}. 
Additionally, these abundant data appear to pose more questions about silicate optical properties than they can currently answer. 
Hence, it is not yet sure how we can best to use these new laboratory data to interpret observations. 

Following some interesting work on annealed iron-containing amorphous silicates \cite{2006A&A...448L...1D}, where it was found that the iron in the silicate is reduced to metal in the presence of carbon it seems reasonable to assume that some (significant) fraction of the cosmic iron is incorporated within amorphous silicates in the form of metallic iron as nano-particle inclusions \cite{2006A&A...448L...1D}. 
In the ISM it is likely that the silicate and carbonaceous dust populations are not completely segregated because some `cross-contamination' must occur. Thus, the interstellar amorphous silicates are almost certainly mixed with a carbonaceous dust component \cite{1989ApJ...341..808M} probably in the form of mantles \cite{2013A&A...558A..62J,1997A&A...323..566L}. Such an intimate mix of carbon and silicate in dust would then provide the ideal  conditions for the reduction of iron into metallic nano-inclusions within the silicate phase. 
The optical properties of an amorphous forsterite-type silicate with iron nano-inclusions are comparable to those for other iron-containing amorphous silicates where the iron is incorporated into the silicate structure \cite{2013A&A...558A..62J}. Thus, and on the basis of the optical properties alone, it is unfortunately not possible to discern the nature of the solid state into which iron is incorporated.  

X-ray absorption and scattering by interstellar dust  \cite{PoSLCDU2013006} 
indicates that a significant fraction of the cosmic iron is not in the form of silicates but rather in a metallic form \cite{2005A&A...444..187C,2011ApJ...738...78X}, consistent with the above supposition about the nature of iron in dust in the ISM. Additionally, early analyses of the handful of extra-solar system grains collected by the Stardust mission indicate the presence of metallic Fe and also seem to hint at the presence of some iron sulphide. 


{\it The bottom line}: is that it is now time to abandon `astronomical' silicate in dust modelling and to be constrained by the laboratory-measured properties of amorphous silicate dust analogues. Unfortunately, these measurements are far from unambiguous. 

Further, there is now very strong evidence from x-ray absorption and scattering data that a large fraction of the cosmic Fe is in a metallic form. 

\section{Dust is the same everywhere}
\label{sect_dust_everywhere}

In the light of the discussions in the preceding sections it seems to be rather evident that interstellar dust must evolve as it transits from region to region, with its properties reflecting its response to its local environment (density, temperature, radiation field, \ldots). Dust models that do not allow for the evolution of the dust composition, structure and optical properties would therefore seem to be of limited use in the interpretation of present day observations ({\it i.e.}, from the {\em Spitzer, Herschel} and {\em Planck} missions). In the following several examples of observations revealing strong dust evolution effects are considered. 

There is now much intriguing evidence for dust evolution in photo-dissociation regions (PDRs), regions where the dust accreted and/or coagulated in dense molecular clouds is exposed to the intense radiation field of a star-forming environment. 
For example, there is evidence for dust evolution in PDRs from the analysis of Spitzer and ISO observations of the Horsehead Nebula and NGC\,2023\,North \cite{2008A&A...491..797C} and Herschel observations of the Orion Bar \cite{2012A&A...541A..19A}, which indicate lower relative abundances of the IR band emitters and the small grains responsible for the mid-IR continuum, by up to an order of magnitude compared to the dust size distribution typical of the diffuse ISM. 
Further, the photo-fragmentation of the small carbon grains, responsible for the mid-IR continuum dust emission,  into nano-particles (PAHs) has been shown to occur in relatively dense, $n_{\rm H} = [10^2, 10^5]$\,cm$^{-3}$, and UV-irradiated PDRs, $G_0 = [10^2, 5 \times10^4]$  \cite{2012A&A...542A..69P} ($G_0$ is the interstellar radiation field intensity in units of the value in the solar neighbourhood), {\it i.e.}, over a relatively narrow range of $G_0/n_{\rm H} = 0.5-1$. 

Other recent work shows that the intensity of the UV extinction bump at 217\,nm does not correlate with carbon depletion into dust or with the FUV extinction and therefore suggests carbonaceous dust evolves even in the neutral ISM \cite{2012ApJ...760...36P}, perhaps through the effects of accretion in denser regions ({\it c.f.}, the discussion in \S \,\ref{sect_dust_spect}) and towards protostars \cite{1994ApJ...433..179S}.  
The carbon depletion is most likely determined by accretion ({\it i.e.}, by the cloud volume along the line of sight), whereas the UV bump intensity and the FUV extinction reflect the abundance of UV photo-processed small grains in the low density surface regions of the clouds. Thus, carbon depletion, the UV bump intensity and the FUV extinction are somewhat de-coupled. 
It was also noted that the FUV extinction appears to show a gradual decrease with decreasing gas density, $n_{\rm H}$, which was interpreted as due to the preferential fragmentation of small grains in the diffuse ISM \cite{2012ApJ...760...36P}. 

There is abundant evidence for dust evolution in the denser regions of the ISM, some of the more recent work in this area, based on Herschel observations of the L1506 dense filament in the Taurus molecular cloud complex \cite{2013A&A...559A.133Y}, indicates an increase in the dust opacity at 250\,$\mu$m by a factor of about two. This result was interpreted as due to grain coagulation into fluffy  aggregates in regions where the gas density is greater than a few $\times 10^3$\,H\,cm$^{-3}$
and for $A_{\rm V} \simeq 2-3$ \cite{2013A&A...559A.133Y}. In contrast, the dust in the cloud outer layers is similar to that in the diffuse ISM. This observational result is consistent with the results of recent dust accretion/coagulation models \cite{2011A&A...528A..96K,2012A&A...548A..61K}. 

{\it The bottom line}: is that dust is far from the same everywhere and that observations clearly show that its properties evolves through photo-fragmentation in energetic PDRs but also by accretion in the diffuse ISM and by accretion and coagulation in dense molecular clouds. 

\section{Concluding remarks}
\label{sect_conclusions}

In the light of the above discussion it seems as though we really do need to take a much closer look at many of our currently-held ideas about dust in the ISM, what it is made of, its structure and its response to its environment. 
So, what do we really know of cosmic dust? 
By way of an incomplete summary here are a few suggestions as to what a re-evaluation of some of our long-held assumptions has led us to today: 
\begin{itemize}

  \item Dust extinction varies, most likely due to important changes in the carbonaceous dust optical properties in the visible region. Thus, extinction must be studied in conjunction with the more diagnostic full dust SED and, ideally, in conjunction with polarisation studies if we are to gain a fuller insight into dust evolution. 

  \item The absolute value and spectral slope of the dust emissivity at FIR-mm wavelengths are {\em not} free parameters and are {\em not} fixed but vary with wavelength, composition and temperature. 

  \item Carbon atoms appear to be more abundant in the ISM than previously assumed and pose no under-abundance problem ({\it a.k.a.}, carbon `crisis') for dust models, except perhaps that the current dust models do not use enough carbon to explain the dust composition in the denser regions of the ISM. 

  \item Dust is primarily composed of two distinct solid phases comprised of different elements, but which can be intimately bound together: amorphous hydrocarbon solids (C, H), Mg-rich amorphous silicates (Mg, Si, O), metallic iron (Fe) and perhaps also iron sulphide (Fe, S). Carbon is probably to be found in isolated grains but also in mantles on amorphous silicate grains. Additionally, most cosmic iron is probably present as metallic nano-inclusions (in an Mg-rich amorphous silicate phase).

  \item The solid phase carbon accreting as mantles onto grains in dense regions of the ISM and towards protostars is probably significantly different from the bulk of the aromatic-rich carbonaceous dust present in the low-density diffuse ISM, in that it probably incorporates a higher atomic hydrogen fraction and is aliphatic-rich.

  \item Carbonaceous dust luminesces in the red and blue. The red luminescence is widespread throughout the ISM but the blue luminescence so far appears to be unique to the Red Rectangle region. These luminescent phenomena should reveal a wealth of critical information about the UV photo-processing of dust in the ISM and circumstellar regions. 

  \item Interstellar carbonaceous dust is not graphite or graphitic but is most likely an amorphous hydrogenated carbon phase, with wide-ranging optical and thermal properties, that can incorporate a significant H atom fraction and possibly other elements as hetero-atoms. 

  \item Dust is {\em not} the same everywhere but undergoes important evolution in response to the local physical conditions (radiation field, density, temperature). PDRs and the outer regions of dense molecular clouds appear to be the most promising environments to study and quantify the processes at the core of dust evolution in the ISM. 

\end{itemize}

In conclusion, it seems that only now and after some eight decades of study are we finally beginning to make some important inroads into advancing our understanding of cosmic dust. Perhaps we are, at last, beginning to peer under the extinguishing veil to glimpse the delights that lie ahead. However, and as always in science, we will need to shoulder the burden of increased complexity if we are to significantly advance beyond our current understanding of cosmic dust. \\

\noindent{\it Acknowledgements.} 
I wish to thank the organisers of the Taiwan dust life cycle meeting for their invaluable help and for a warm welcome in Taipei. I also wish to thank many colleagues for the innumerable and invaluable discussions over the years that have helped me to distill my ideas into the views presented in this review. \\

\noindent{\it Disclaimer.} 
The views in this review reflect the interests and biases of the author. The reader is therefore advised to regard or disregard them as her/his own personal biases dictate. However, it is the author's hope that the ideas discussed here might help to direct future cosmic dust studies. 

\bibliography{biblio_HAC}
\bibliographystyle{JHEP}

%

%
%
%

\end{document}